# Comparing with Python: Text Analysis in Stata

**Xiangtai Zuo (Shutter Zor)**
ORCID: 0000-0003-1286-4765
Google Scholar: https://scholar.google.com/citations?user=IWnqLLIAAAAJ
School of Management, Xiamen University, Xiamen, Fujian, China
Shutter_Z@outlook.com

**Abstract**
Text analysis is the process of constructing structured data from unstructured textual content, usually implemented in Python. In terms of the principles of text analysis, a computer program with the ability to read a file and match it with a regular expression is all that is needed for basic text analysis. However, few researchers have used Stata as their main text analysis tool. In this paper, I will take a step-by-step approach to the practical process, giving examples of how text analysis can be performed with Stata, and comparing the code and running time with Python.

## 1 Introduction

Text Analysis is the application of NLP (Natural Language Processing) to textual content for automated information extraction or measurement (Bochkay et al., 2023). Since the turn of the century, text analysis has been increasingly used in social science research (Schwartz and Ungar, 2015; Gentzkow et al., 2019; Berger et al., 2020; Colón-Ruiz and Segura-Bedmar, 2020; Loughran and McDonald, 2020). Python was still the most frequently mentioned software in these extensive and useful studies. However, for most researchers in economics and management, Stata is the most widely used econometric software because of its adequate statistical functionality and repeatability (MacKie-Mason, 1992; Renfro, 2004; Greene, 2015). It is worth noting that Stata, despite being in its 18th version, still lacks an official command for the relevant text analysis methods. Therefore, when researchers in economics and management want to use textual data to support their interesting conclusions, Python becomes an inescapable hurdle.

     However, learning comes at a cost in time and effort, especially when learning new tools. To reduce the learning costs for researchers and to address the difficulties of text analysis of long papers in Stata, particularly word frequency statistics. This paper proposes a convenient method for counting word frequencies using Stata and expects to contribute to the long-term development of Stata.

## 2 Overview of text analysis

The basic building blocks of text are words and, as mentioned above, a specific application of NLP (Bochkay et al., 2023). Therefore, text analysis cannot be separated from counting the words in the content. Using the most basic method of counting the number of occurrences of each word in a text, it is possible to calculate, for example, the tone (or sentiment) of the text (Loughran and McDonald, 2011), economic policy uncertainty (Baker et al., 2016; Wang et al., 2023), digital transformation (Tu and He, 2022; Zhu et al., 2022), etc. Furthermore, a count-based vector or a binary-based vector can be constructed depending on the number of words or whether they occur. Different similarity metrics can then be constructed based on these vectors with different degrees of weighting. These metrics are widely used in the calculation of conservative and revolutionary innovations (Zor, 2023).

Counting word occurrences in text forms the basis for other text analyses. Admittedly, there are several third-party text analysis packages available for download from Stata, e.g. txttool (Williams and Williams, 2014), wordfreq (Dicle and Dicle, 2018), ldagibbs (Schwarz, 2018), and lsemantica (Schwarz, 2019), etc. However, the lack of direct comparison with Python makes it difficult for Stata users to determine whether they need to re-learn Python when working on text analysis. Therefore, in the next section, I will describe how word frequency statistics are implemented in Stata and Python, respectively, and compare the code and performance.

## 2 Word frequency statistics in Stata and Python

As the software and running time comparison are covered in the following text analysis implementation, the software version and environment are described first. The system on my computer is Windows 11 and it is running with 16G of RAM. In the following tests, I use version "MP 16.0" of Stata and version "3.8.13" of Python. Also, to avoid unnecessary functions in third-party programs that take up space at runtime, I only use Stata's (Python's) default (standard) functions to implement word frequency statistics.

Next, I will show code for text analysis (word frequency statistics) using Stata and Python respectively, and make some comparisons. The text I am using is the Shakespeare book *The Tragedy of Hamlet, Prince of Denmark*.

### 2.1 Comparison of reading txt files

In Stata, I use the `import` to read the information from the txt file. In Python, I am using the `with open` to read the information from a txt file. Table 1 shows the comparison of code and performance.

Table 1. Comparing code and performance when reading a txt file

| Software | Stata | Python |
|---|---|---|
| **Code** | import delimited "TestFile.txt", delimiter("shutterzor", asstring) varnames(nonames) clear | with open("TestFile.txt", "r") as file:<br>    Text = file.read() |
| **Time*** | 0.017s | 0.003s |
| **Result** | A new data frame | A new variable |

Note: *, Average time is taken after three measurements.

When Stata reads into a txt file, I use a long string (shutterzor) that is unlikely to appear in the text, to ensure that there is only one variable in the data window. Stata imports data and displays it in the data window, whereas in Python it is stored as a variable (similar to a pane in the Stata data window). This is a slight difference between Stata and Python when working with text data, and will also make a difference to the specific analysis methods that will follow. In terms of performance, Stata generated a variable with 5382 observations in 0.017s, while Python generated only one observation (variable) in 0.003s.

2.2 Comparison of lowercase conversions

Once the text data has been read in, the text analysis process requires all letters to be converted to lowercase. Table 2 shows the code and performance comparison for the lowercase conversion.

Table 2. Comparing code and performance when lowercasing letters

| Software | Stata | Python |
|---|---|---|
| Code | replace Text = lower(Text) | Text = Text.lower() |
| Time* | 0.002s | 0.000s |

Note: * Average time is taken after three measurements.

When it comes to converting letters to lowercase, the performance of these two programs is comparable.

2.3 Comparison of removing punctuations

Some extra punctuation, tabs ("\t"), and line feeds ("\n") have no real meaning, carry less information, and can affect the results of text analysis, so it is generally necessary to remove them before calculating word frequency. Table 3 shows the code and performance for excluding text punctuation in Stata and Python respectively.

Table 3. Comparing code and performance when removing punctuations

| Software | Stata | Python |
|---|---|---|
| Code | local punctuation "! # $ % & ' ( ) * + , - . / : ; < = > ? @ [ \ ] ^ _ { | } ~ " `" <br><br> foreach marks of local punctuation{ <br>     replace Text = subinstr(Text, "`marks'", " ", .) <br> } <br><br> replace Text = ustrregexra(Text, "\t", " ") <br> replace Text = ustrregexra(Text, "\n", " ") | Text = Text.replace("'", " ") <br><br> for punctuation in '!"#$%&()*+,-./:;<=>?@[\\]^_`{|}~': <br>     Text = Text.replace(punctuation, " ") <br><br> Text = Text.replace("\n", " ") <br> Text = Text.replace("\t", " ") |
| Time* | 0.055s | 0.003s |

Note: * Average time is taken after three measurements.

Python is better than Stata at removing punctuation, but there is no way for Python to put a single quote directly between two single quotes, or a single double quote between two double quotes when doing a loop without throwing an error. Stata's local, however, handles this situation very well.

2.4 Comparison of word frequency calculations

By doing the above, the txt file we read has now become a variable consisting only of lowercase words with spaces, as shown in Figure 1.1 and Figure 1.2.

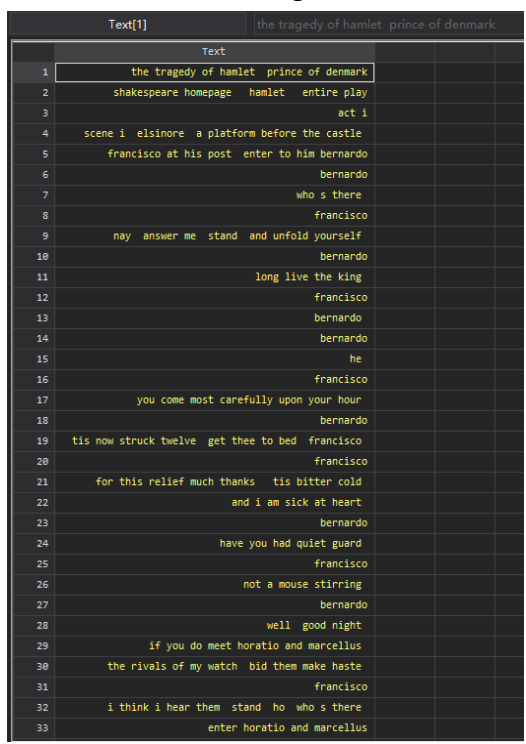
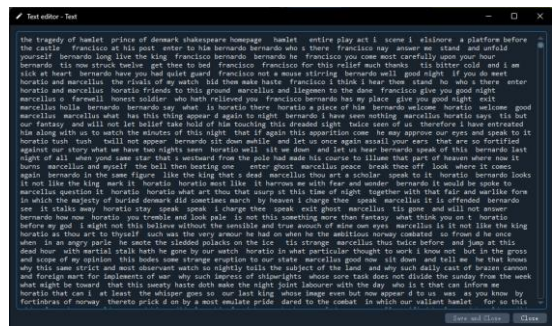

Figure 1.1 Text in Stata          Figure 1.2 Text in Python

Using these two graphs, it is possible to visualize the differences between Stata and Python after reading in the data mentioned in subsection 2.1. While it is possible to put all this text into a single box in Stata, this makes it difficult to take advantage of Stata's data-handling capabilities. Moreover, if all of these observations were combined into a single observation, text segmentation using `tokenize` would generate a large number of duplicate variables for recording word frequency, which would easily exceed Stata's default processing limits.

In the end, I found a method that I thought was more suitable for use with Stata's word frequency statistics. That is, following the form of variable storage in Figure 1.1, each observation would be stored in a different `tempfile`, and then the word frequencies in the `tempfile` would be calculated using `egen` and `count()`, and finally the statistics would be combined. And in Python, word frequency will be counted using the dictionary method. Table 4 shows the code and performance for calculating word frequency in Stata and Python respectively.

Table 4. Comparing code and performance when calculating word frequency

| Software | Stata | Python |
|---|---|---|
| **Code** | split Text, parse(" ") <br> drop Text <br> local Num = 1 <br> foreach variable of varlist _all { <br>     tempfile file`Num' <br>     preserve <br>       keep `variable' <br>       bys `variable': egen Count = count(`variable') | Words = Text.split() <br> Counts = {} <br> for word in Words: <br>     Counts[word] = Counts.get(word, 0) + 1 <br><br> Items = list(Counts.items()) <br> Items.sort(key = lambda x:x[0]) |

```
        rename `variable' Word
        duplicates drop Word, force
        save "`file`Num'"
    restore
    local Num = `Num' + 1
}

clear
gen Word = ""
gen Count = .

local fileNum = `Num' - 1
forvalues i = 1/`fileNum' {
    append using "`file`i''"
}
drop if Word == ""
bys Word: egen Total = sum(Count)
duplicates drop Word, force
keep Word Total
```

| Time* | 2.538s | 0.015s |
|---|---|---|

Note: * Average time is taken after three measurements.

Stata took longer but processed 5382 files (observations) at the same time because each observation was processed into a single file. Python took less time but only processed one variable because there was only one observation (variable) in the RAM.

Finally, the word frequency results for Stata and Python are the same.

### 3 Conclusions

As textual information becomes increasingly important to researchers, more and more text analysis methods are being discovered and learned. However, mainstream text analysis software is still dominated by Python. Stata, a tool used by a wide range of researchers, has received less attention for its text analysis capabilities.

Based on this real-world context, this paper compares the capabilities of Stata and Python in text analysis, specifically word frequency statistics, both in terms of code and performance. I argue that with native function support, Stata's text analysis performance is no worse than Python's, and may even be better.

Finally, based on the technical conclusions of this article and my experience using Stata, I would suggest that Stata technicians consider enhancements to the macro function in future releases, such as introducing dictionary-type macro settings. Also, when designing the program, I asked ChatGPT to try to get information about word frequency statistics from Stata. It gave me a macro called `dictionarize` and wanted me to store word frequency data through this macro, as Python does. The truth is that Stata does not have such a macro. Therefore, both from my practice and from ChatGPT's answers, we believe that Stata should include something like a dictionary macro for longer term development.

**Appendix**

1. The word frequency statistics function in this article can be downloaded by typing "ssc install onetext" in Stata, a command that I have wrapped.

2. The data and code used in this article can be decompressed from the attached file.